\documentclass[pdflatex,sn-mathphys-num]{sn-jnl}
\usepackage{graphicx}%
\usepackage{multirow}%
\usepackage{amsmath,amssymb,amsfonts}%
\usepackage{amsthm}%
\usepackage{mathrsfs}%
\usepackage[title]{appendix}%
\usepackage{xcolor}%
\usepackage{textcomp}%
\usepackage{manyfoot}%
\usepackage{booktabs}%
\usepackage{algorithm}%
\usepackage{algorithmicx}%
\usepackage{algpseudocode}%
\usepackage{listings}%

\theoremstyle{thmstyleone}%
\theoremstyle{thmstyletwo}%

\theoremstyle{thmstylethree}%

\begin{document}

\title[Article Title]{Spin-Split Dispersion of Leaky Surface plasmons in Inversion- Symmetric System}


\author[1]{\fnm{Sujit Rajak}} 
\equalcont{These authors contributed equally to this work.}

\author[2]{\fnm{Nishkarsh Kumar} }
\equalcont{These authors contributed equally to this work.}

\author[1]{\fnm{Dheeraj Yadav}} 

\author[1]{\fnm{Suman Mandal} }
\author[3]{\fnm{Jeeban K. Nayak} }
\author[1]{\fnm{Ayan Banerjee} }
\author[1,4] {\fnm{Subhasish Dutta Gupta} }
\author[3]{\fnm{Olivier Martin} }
\author[1]{\fnm{Nirmalya Ghosh} }

\affil[1]{\orgdiv{Department of Physical Science}, \orgname{Indian Institute of Science Education and Research Kolkata}, \orgaddress{ \city{Mohanpur}, \postcode{741246}, \state{W.B.}, \country{India}}}

\affil[2]{\orgdiv{Department of Physics}, \orgname{Indian Institute of Technology Kanpur}, \orgaddress{\city{Kanpur}, \postcode{208016}, \state{U.P.}, \country{India}}}

\affil[3]{\orgdiv{Nanophotonics and Metrology Laboratory (NAM)}, \orgname{ Swiss Federal Institute of Technology Lausanne (EPFL)}, \orgaddress{\city{Lausanne},  \postcode{1015}, \country{Switzerland}}}

\affil[4]{\orgdiv{Tata Centre for Interdisciplinary Sciences}, \orgname{ TIFRH}, \orgaddress{ \city{Hyderabad}, \postcode{ 500107}, \state{Telangana}, \country{India}}}


\abstract{Spin-dependent dispersion and Rashba effect are manifestations of universal spin orbit interaction (SOI) associated with the breaking of the spatial inversion symmetry in condensed matter and in optical systems. In sharp contrast to this,  we report a spin-split dispersion effect of leaky surface plasmons  in an inversion-symmetric one dimensional plasmonic grating system.  In our system, the signature of spin-momentum locking and the resulting spin-polarization dependent splitting of dispersion of the surface plasmons are observed through the leakage radiation detected in a Fourier (momentum) domain optical arrangement. The setup enables single-shot recording of the full polarization-resolved dispersion  (frequency $\omega$ vs transverse momentum $\mathbf{k}$) of the leaky surface plasmons. Momentum domain polarization  analysis  identified a transverse momentum ($\mathbf{k}$)-dependent linear birefringence-linear dichroism effect (referred to as the “geometric LB-LD effect”) responsible for the observed spin-split dispersion. This unconventional SOI effect is reminiscent of the recently reported LB-LD effect resulting in giant chirality in centrosymmetric crystal, albeit with geometric origin. It is demonstrated that the interplay of the geometrical polarization transformation in focused polarized light and  subsequent interaction of the structured field polarization with the plasmonic grating leads to the evolution of strong geometrical phase gradient or spin(circular polarization)-dependent transverse momentum of light resulting in spin-split dispersion. Our study offers a new paradigm of spin-based dispersion engineering and spin-enabled nano-optical functionalities in simple symmetric metasurfaces using geometric LB-LD effect.   }

\keywords{}



\maketitle

\section{Introduction}\label{sec1}
Spin orbit interaction (SOI) refers to the relativistic interaction of a particle’s spin degree of freedom with its orbital degree of freedom. The SOI is a universal phenomenon observed in diverse fields of physics,  spanning from atomic, condensed-matter systems to optical systems\cite{soumyanarayanan2016emergent, Bliokh2015SOI}. Spin orbit coupling in condensed matter systems  leads to a variety of intriguing transport phenomena like Spin Hall effect (SHE), Spin-momentum locking, Rashba effect etc \cite{wunderlich2005experimental,zhang2005intrinsic, kato2004observation, sanchez2013spin}. The  Rashba effect in the broken inversion symmetric system that leads to the splitting of spin degenerate parabolic electronic bands into bands of opposite spins, has attracted particular attention due to its fundamental nature and potential applications \cite{Varotto2022Rashba,aiello2015,shitrit2013spin,PhysRevLett.105.136402}. Since light carries both spin angular momentum (SAM – associated with circular / elliptical polarization) and orbital angular momentum (OAM – associated with helical phase front of light beams), coupling between these two degrees of freedom in various light-matter interactions have  led to analogous optical SOI phenomena like the SHE of light, optical Magnus effect, plasmonic Ahronov-Bohm effect, geometrical Doppler effect and optical Rashba effect etc\cite{Bliokh2015SOI,Shao2018TransverseSOI}. Evolution of geometric phase and its spatial (or momentum) gradient is at the heart of most of the optical SOI phenomena \cite{Bliokh2015SOI, Shao2018TransverseSOI,Neugebauer2013GeometricSH,DuttaGupta2015WaveOptics}. The SHE of light and the related SOI phenomena originating from the spatial gradient of Pancharatnam Berry (PB) phase in spatially inhomogeneous anisotropic medium (the so-called inversion symmetry-broken metasurfaces) are being actively investigated due to potential spin-controlled nanophotonic applications \cite{Xie2021PBMeta,marucci2006,kildishev2013}.  Large PB geometric phase gradient in these nano-structured metasurfaces comprising of anisotropic optical nano-antennas with spatially varying anisotropy axis orientation $\beta(x,y)$ leads to the emergence of giant SHE of light\cite{Xie2021PBMeta,Yin2013SpinHall}, which have enabled dispersion engineering in spin-dependent manner as in the case of Rashba effect in solids \cite{Yin2013SpinHall,liu2015,shitrit2013spin,PhysRevLett.105.136402,rodriguez2013near}. Similar to the role of the potential gradient that breaks the spatial inversion symmetry in the electronic Rashba effect, the space-variant orientation angle of the anisotropy axis  $\beta(x,y)$ induces a spin-split dispersion of optical modes through an additional SAM-dependent transverse momentum $k_g=\sigma \nabla \beta$, where $\sigma=\pm1$ for right and left circular polarizations, $\nabla \beta $ accounts for the spatial gradient of PB phase\cite{Xie2021PBMeta,Yin2013SpinHall,shitrit2013spin,PhysRevLett.105.136402,rodriguez2013near}.  Drawing parallels with spintronics, a variety of symmetry-broken metasurfaces have been designed through complex nano-structuring to obtain diverse types of spin-controlled optical functionalities, which has opened novel route towards development of next generation spin-controlled  nanophotonic devices for diverse applications ranging from integrated photonic circuits,
sensing and metrology, microscopy, quantum
information processing and so forth.\cite{wunderlich2005experimental,zhang2005intrinsic, kato2004observation, sanchez2013spin,Yu2014FlatOptics,kildishev2013,aiello2015}.  In this regard, alternative approaches that may enable similar spin-optical functionalities using relatively simpler metasurfaces (having relative ease of fabrication) are highly sought after.
 \cite{DorrahCapasso2021TunableStructuredLight}. \\

Here we report a remarkable spin polarization-dependent dispersion effect of leaky surface plasmons in a simple one-dimensional plasmonic crystal (comprised of gold grating)  exhibiting perfect spatial inversion symmetry. The spin (circular polarization)-dependent dispersion of the surface plasmons is observed by performing polarization and spectrally-resolved detection of the leakage radiation or the scattered light from the plasmonic crystal in the momentum ($\mathbf{k}$) domain using a Fourier optical arrangement\cite{PhysRevLett.131.193803,lin2013polarization,rodriguez2013near}. A  momentum domain polarization Mueller matrix experimental system enabled us to simultaneously detect the polarization resolved transverse momentum ($k_{x/y}$) distribution of the leaky modes in the far-field and the polarization resolved dispersion ($E=\hbar\omega \  vs \ k$) features  \cite{kvasnivcka2014convenient}\cite{drezet2008leakage}. The spin-split dispersion of the leaky surface plasmons are manifested in the detected circular polarization-resolved intensities as a consequence of spin-momentum locking\cite{PhysRevLett.131.193803}. Theoretical modeling and analysis of the geometric polarization evolution of tightly focused light through the anisotropic plasmonic grating unraveled the origin of such giant chiral effect in the  inversion-symmetric system as a momentum domain  “geometric” linear birefringence-linear dichroism (LB-LD) effect. This effect originates from the interplay of linear birefringence (retardance) and linear dichroism (or vice versa) having different orientation of optical anisotropy axes. Recently, an analogous LB-LD effect associated with "intrinsic" linear anisotropy of the system has been shown to cause giant circular dichroism effect (differential absorption of circularly polarized light) in centrosymmetric crystal \cite{Parrish2025CentrosymmetricCD}, which is in contrary to the usual perception of structure-property relations. The giant spin (circular polarization)-dependent optical effect that we observe here in the inversion symmetric plasmonic crystal system is fundamentally different and originates SOI of light and evolution of geometric phase which is completely governed by the geometry of polarization evolution. This therefore manifests as a “geometric” LB-LD effect that depends on the trajectory of the polarization evolution, i.e., on the transverse momentum ($k_{x/y}$) of the scattered light. As a consequence of this momentum-domain geometrodynamics of polarized light, a strong  geometric phase gradient is generated which leads to spin-dependent transverse momentum leading to the spin-split dispersion like in Rashba effect. We have subsequently quantified this spin-dependent transverse momentum using the momentum matching conditions for the two-step process of the excitation of the surface plasmons and subsequent emission of the leakage radiation. To the best of our knowledge, this is the first experimental evidence of spin-split dispersion of optical modes in  inversion-symmetric  system induced by the “geometric” LB-LD effect. \\
\section{Results and Discussion}\label{sec2}
\begin{figure}
    \centering
    \includegraphics[width=1\linewidth]{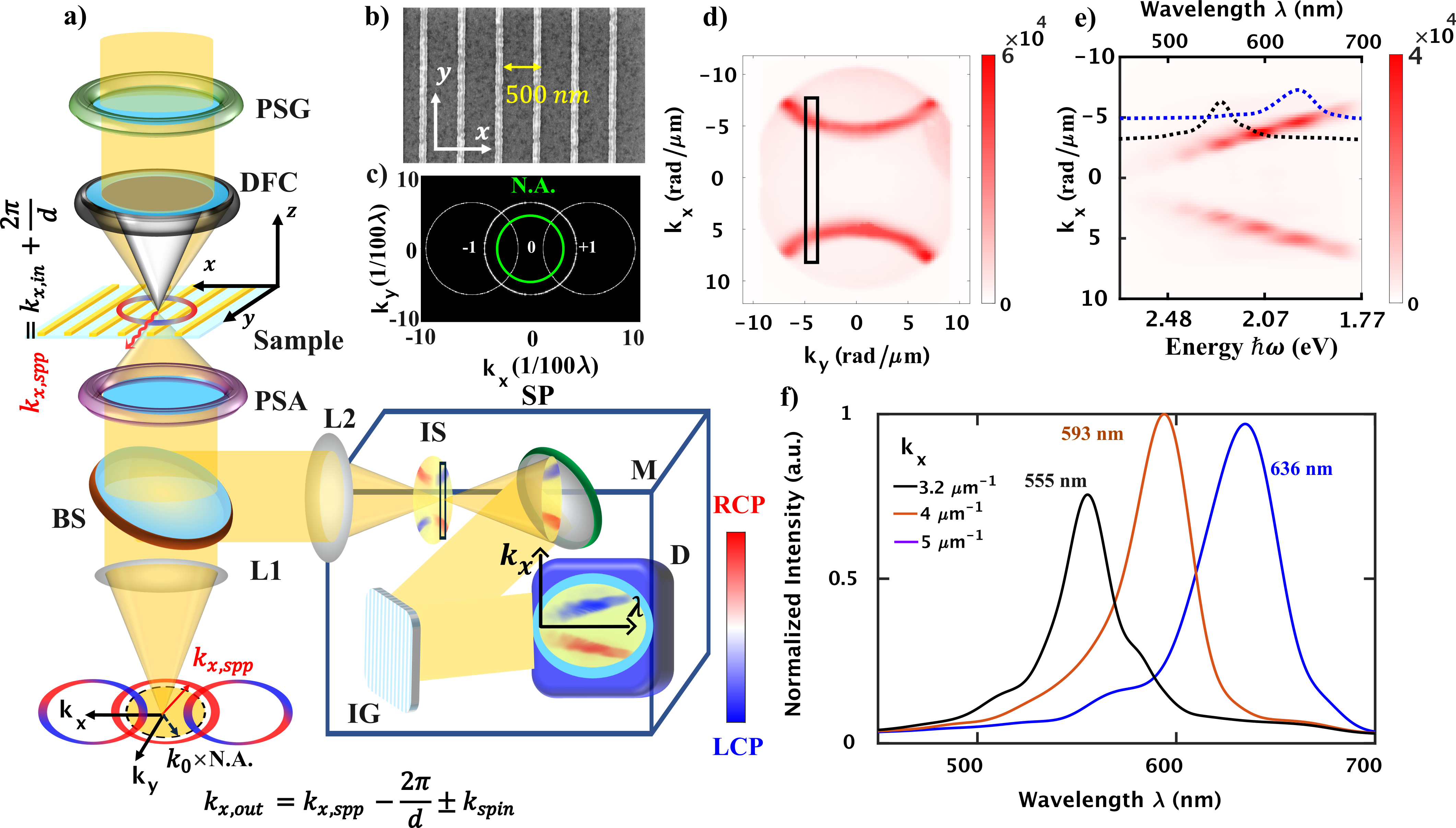}
    \caption{\textbf{Experimental embodiment for recording SOI of light in the momentum ($k_x,k_y$) domain
and for the detection of the polarization-resolved dispersion ($E/\lambda$ vs $k_x$) of the surface plasmon
modes of the plasmonic grating  through the leakage radiation or scattered light in
the far-field Fourier domain:}(a) Schematic of the experimental setup, PSG, DFC (dark field condenser), PSA, L1 and L2 (lenses), M (mirror), BS (beam splitter), and SP (spectrograph). (b) Scanning electron microscope (SEM) image of the 1D plasmonic grating with a period of $500 nm$. (c) Simulated and (d) the experimentally recorded polarization-blind momentum domain intensity arc segments of the diffraction pattern  formed by the leaky surface plasmons. (e) Dispersion characteristics ($k_x$ vs. $E/\lambda$) is recorded from the region marked by the black rectangle in (d). The bottom axis shows the energy ($E=\hbar \omega$) and the corresponding wavelength ($\lambda$) is shown in the top axis. (f) Spectral response obtained at different transverse momentum values ($k_x$) along the dispersion curve, showing the plasmon resonances.}
    \label{fig:1}
\end{figure}

\subsection*{\textit{Experimental observation of spin-split dispersion from inversion symmetric 1D plasmonic crystal:}}

We have employed a dark-field microscopic system to record transverse momentum ($k_{x/y}$)-resolved polarization response of the  plasmonic grating, as shown in Fig. \ref{fig:1}(a)\cite{PhysRevLett.131.193803}. The set-up utilizes broadband white light illumination, and incorporates PSG (polarization state generator) and PSA (polarization state analyzer) units in the path to capture the polarization-resolved ($k_{x/y}$) distributions of the scattered light (leakage radiation) from the surface plasmons  for the wavelength range of 400–700 nm.\cite{kvasnivcka2014convenient}\cite{drezet2008leakage}. By sequentially generating and analyzing six independent polarization states (linear horizontal, vertical, $\pm45^\circ$, and right- and left-handed circular), the system enables the construction of full $k$-domain $4\times4$ Mueller matrix ($M(k_x,k_y)$) of the sample, which contain complete information on the nature of the polarization transformation (see Supporting information S1). The 1D plasmonic crystal consists of gold gratings with a period ($d$) of $500 \ \text{nm}$, which is fabricated (see Supporting information S2) on a quartz substrate using electron-beam lithography (SEM image shown in Fig. \ref{fig:1}(b)). The excitation of surface plasmons (SP) in the gold grating is governed by the momentum matching condition: $k_{spp} =k_{x,in} + \frac{2\pi}{d}$, where $k_{spp}$ is the in-plane momentum of the SP mode, $k_{x,in}$ is the transverse momentum of the incident light\cite{loren2023microscopic}. In our setup, this condition is fulfilled by employing a dark-field condenser with a high numerical aperture (NA = 0.8–0.92), which allows access to large transverse momentum components ($k_{x}, k_{y}$). The transverse momentum $k_{x}$ (parallel to the grating vector) enables coupling of the incident light into the leaky surface plasmons at the metal–dielectric interface, whose radiation is out-coupled through the periodic grating and collected in the far-field with a high-NA objective\cite{lopez2013fundaments}. This leakage radiation forms diffraction rings in the Fourier plane \cite{lopez2013fundaments,kvasnivcka2014convenient}(see Supporting information S3). As illustrated in Fig. \ref{fig:1}(c), the radius of the diffraction ring  represents the surface plasmon wavevector $k_{spp}$, and the distance between the center of adjacent rings is given by $2\pi/d$. Note that the zeroth-order diffraction ring is not captured in our dark-field microscopic arrangement, and only intensity arc segments of the shifted \(+1\) and \(-1\) diffraction orders are recorded, as the accessible spatial frequency (\(k\)) is limited by the NA of the objective ($k_{spp}>2\pi (\text{N.A.})/\lambda$ as illustrated in Fig. \ref{fig:1}(d) (shown for the spectral band $\lambda = 610-635 \ \text{nm}$ recorded using a band pass filter).

In order to record the dispersion characteristics of the leaky surface plasmons, the input slit of a spectrometer is positioned at the Fourier plane aligned along the $k_{x}$ direction ( Fig.\ref{fig:1}(d)). The spectrometer then captures $k_{x,out}$-resolved spectral intensity profiles $I(\lambda)$ for the wavelength range $450$–$700\ \text{nm}$. For this purpose, the Fourier plane coordinates at the entrance slit were calibrated for the transverse momentum of the scattered light $k_{x,out}$.\cite{PhysRevLett.105.136402,zhang2020controlling}. The recorded dispersion characteristics of the leaky surface plasmons are shown in Fig. \ref{fig:1}(e). The leakage radiation from the surface plasmons satisfies the momentum matching condition: $k_{x,out} =k_{x,in} \pm k_g^{x}=k_{spp} - \frac{2\pi}{d} \pm k_g^{x}$ \cite{loren2023microscopic}, where $k_g^{x}$ is the x-component of the additional spin-dependent transverse momentum  ($+$ for right circular polarization - RCP and $-$ for left circular polarization-LCP) originating from SOI, discussed subsequently). The recorded $k_{x,out}$-resolved spectra ( Fig.\ref{fig:1}(f)) show distinct intensity peaks as signature of the  leaky surface plasmon resonances, with peaks at 636 nm (blue), 593 nm (brown), and 555 nm (black) wavelengths corresponding to $k_{x,out}$ of 5, 4, and 3.2 $rad/\mu m$, respectively. Once these surface plasmon resonances  are identified, the polarization-resolved dispersion measurements are performed in the form of the $4\times4$ Mueller matrix, $M(E=\hbar \omega$,$k_{x,out})$.

\begin{figure}
    \centering
    \includegraphics[width=1\linewidth]{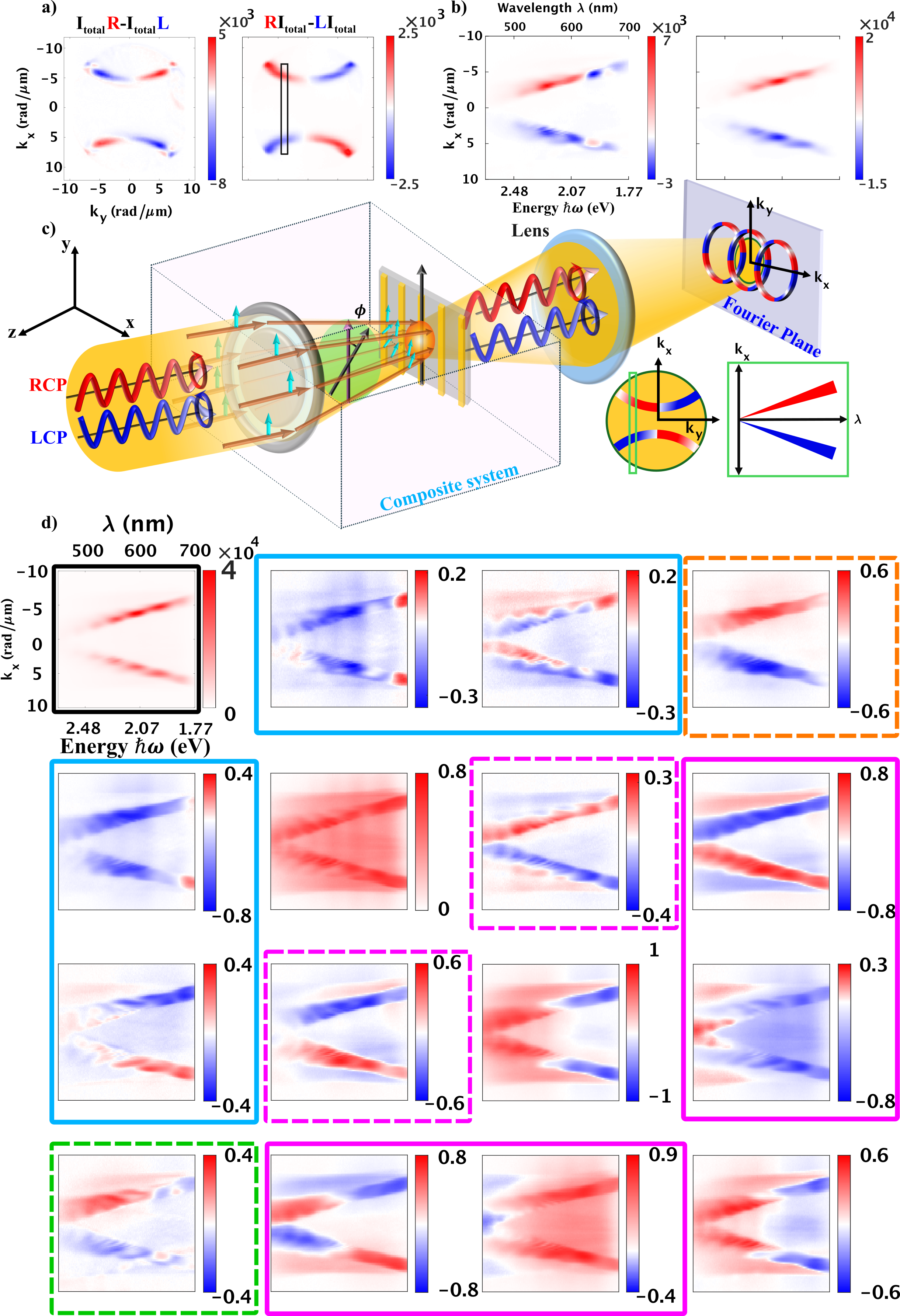}
    \caption{\textbf{Manifestation of spin-split dispersion of the leaky surface plasmons and Spin
momentum locking in the recorded polarization-resolved dispersion features and in the
momentum domain Mueller matrix images:} 
(a) Spin-dependent azimuthal ((\( \phi \))) intensity lobes in the recorded polarization-resolved momentum domain intensity arc segments of the leakage radiation for the spectral band $\lambda = 610$–$635 \ \text{nm}$. Left: Differential circular polarization response under polarization-blind or unpolarized illumination (corresponding to \( M_{41} \) Mueller matrix element). Right: Intensity difference between RCP and LCP (corresponding to \( M_{14} \) element).  (b) The corresponding spin-dependent dispersion across $\lambda = 450$–$700 \ \text{nm}$ are shown as circular polarization response differences (left) and the splitting of RCP and LCP branches (right).
(c) Schematic illustration of the mechanism of the underlying geometric LB–LD effect. The two-step polarization evolution of tightly focused light and its propagation through the anisotropic plasmonic grating are illustrated as two sequential linear diattenuating retarder polarization transformations with different relative orientation of anisotropy axes. The azimuthal (\( \phi \)) orientation of the anisotropy axis of focusing yields \( \phi \)-dependent relative orientation with respect to the fixed anisotropy axis of the plasmonic grating, leading to the geometric LB-LD effect.
(d) Polarization-resolved dispersion plots ($k_{x}$ vs. $E/\lambda$) in the form of Mueller matrix. $M_{11}$ shows the polarization-blind dispersion. Linear diattenuation and linear (circular) retardance descriptor elements are highlighted with blue solid and magenta solid (dashed) boxes. Circular anisotropy elements $M_{14}$ (orange dashed box) and $M_{41}$ (green dashed box) reveal spin-dependent dispersion for LCP ($\sigma = -1$) and RCP ($\sigma = +1$) states.}
    \label{fig:2}
\end{figure}

We have separately recorded and analyzed the momentum domain Mueller matrix images \( M(k_{x}, k_{y}) \) corresponding to the Fourier domain intensity arc segments from the plasmonic crystal (shown in Supporting Fig.S1 for the spectral band 610–635 nm). The ($\mathbf{k}$) domain Mueller matrix encodes characteristic features of different contributing SOI effects of inhomogeneous (spatially varying) anisotropic media in the form of four-fold azimuthal (\( \phi \)) intensity lobes in different off-diagonal elements, which contain complete information on polarization transformation (linear to circular/elliptical or vice versa) associated with SOI. The four-fold azimuthal intensity lobes are well known signatures of azimuthal geometric phase that mediates SOI in vortex linear diattenuator (reflected in \( M_{12/21} \), \( M_{13/31} \) elements) and vortex linear retarder (\( M_{24/42} \) , \( M_{34/43} \) ) systems having azimuthally varying orientation of anisotropy axis\cite{PhysRevLett.131.193803,DuttaGupta2015WaveOptics,bliokh2015spin}.  Importantly, the circular anisotropy descriptor  \( M_{14} \) and \( M_{41} \) elements  exhibited distinct azimuthal intensity patterns oriented along orthogonal radial directions (Fig. \ref{fig:2}(a)). Note that while \( M_{14} \) quantifies circular diattenuation (differential scattering response between RCP and LCP input states), \( M_{41} \) encodes circular polarizance (differential scattering between RCP and LCP output states for unpolarized input). The observed momentum domain features of \( M_{14} \) and \( M_{41} \) are indicative of spin–momentum locking, wherein the transverse momentum of the scattered field is coupled to the SAM (\( \sigma = \pm1 \)) of incident light through SOI. These can also be interpreted as far-field manifestation of SHE of light.  

The polarization-resolved dispersion (\(E\) vs. \(k_{x,out}\)) of the leaky surface plasmons across 450–700 nm wavelength range are shown in Fig. \ref{fig:2}(a). The dispersion profiles (\(E\) vs. \(k_{x,out}\)) recorded under polarization-blind (unpolarized) excitation exhibit clear spin-dependent splitting while analyzed through left (\(\sigma = -1\))- and right (\(\sigma = +1\))-circular polarizations, which yields distinct spectral branches centered at opposite (positive and negative, respectively) transverse momenta (\(k_{x,out}\)) (left panel of Fig.\ref{fig:2}(b)). Analogously, when the system is illuminated with RCP/LCP light (\(\sigma = \pm 1\)), the total scattered intensity exhibits two well-separated dispersion branches (right panel of Fig. \ref{fig:2}(b)). This spin (circular polarization)-dependent dispersion of the leaky surface plasmons  implies a photonic Rashba effect \cite{Parrish2025CentrosymmetricCD,liu2015,aiello2015}. While such effects have been observed in systems having broken inversion symmetry\cite{Parrish2025CentrosymmetricCD}, their emergence in an inversion-symmetric  system is quite unusual. In what follows, (a) we present  experimental Mueller matrix dispersion  $M(E,k_{x,out})$ and Mueller matrix modeling of the polarization evolution in our system to unravel the origin of such unconventional SOI effect as the "geometric" LB-LD effect, (b) we provide  evidence of the geometric nature of the LB-LD effect by analyzing the transverse momentum ($k_{x,out}$)-integrated and $k_{x,out}$-resolved spectral Mueller matrices $M(\lambda)$, and (c) we demonstrate how the azimuthal gradient of geometric phase generated via the geometric LB-LD effect leads to the spin-dependent dispersion in the inversion symmetric plasmonic grating system.  

\subsection*{\textit{Unraveling the geometric origin of the LB-LD effect responsible for  spin-split dispersion in the inversion symmetric system:}}
 The polarization evolution in the two-step process of tight focusing and its subsequent interaction with the anisotropic plasmonic grating can be modeled using sequential product of two Mueller-Jones matrices (see Supporting information S4). Each of the  Mueller matrix represents a linear diattenuating retarder $M(d,\delta,\theta)$ in the momentum domain, exhibiting both linear reatardance or linear birefringence (LB) and linear dichroism (LD) effects\cite{DuttaGupta2015WaveOptics,bliokh2015spin,marucci2006,liberman1992spin}. Here, the first matrix, \( M(d_f, \delta_f, \phi) \), represents the geometrical polarization transformation induced by focusing. Due to the vectorial nature of the focusing transformation, the  anisotropy axis ($\theta$) rotates with the azimuthal angle \( \phi \) in the momentum domain. In contrast, the second matrix, \( M(d_p, \delta_p, 0) \), represents the intrinsic linear anisotropic response of the plasmonic grating, with $\theta$ fixed along the direction of the grating vector ($\theta=0^o$). 
 The relative orientation of the anisotropy axes (\( \phi \)) between the two sequential polarization effects  leads to the so-called LB-LD effect.  The resulting Mueller matrix exhibits non-zero circular anisotropy elements \( M_{14}(k) \) and \( M_{41}(k) \) (see Eq. 6 in Supporting information S4), even in the absence of intrinsic chirality of the system.
 Specifically, the \(M_{14}\) element originates from the LB effect of the tightly focused polarized field together with the LD effect of the plasmonic-crystal, and the \(M_{41}\) element results from the LD effect of the tightly focused polarized light combined with the LB effect of the plasmonic-crystal. The effect is maximum for \( \phi = 45^\circ\) and vanishes for \( \phi = 0^\circ\) or \(90^\circ\). This explains the observed circular polarization (spin)-dependent response from the plasmonic grating having perfect inversion symmetry and no intrinsic chirality (illustrated in Fig. \ref{fig:2}c). The resemblance and the fundamental differences between this effect and the extraordinary circular dichroism in centrosymmetric crystals arising from  "intrinsic"  LB-LD effect \cite{Parrish2025CentrosymmetricCD,meng2025dynamic} worth a brief mention. Unlike the intrinsic LB-LD effect, here, the observed circular anisotropy  depends upon the transverse momentum  of light ($k_{x/y}$) (or on the trajectory of light) \cite{bliokh2015spin,DuttaGupta2015WaveOptics,marucci2006,liberman1992spin} and is governed by the geometry rather than the material symmetry.  The recorded polarization-resolved dispersion \( M(E, \mathbf{k}_{x,\text{out}}) \) in the form of Mueller matrix are presented in (Fig. \ref{fig:2}(d)).  The presence of all non-zero off-diagonal elements illustrate the complex nature of polarization transformation as a consequence of the geometric LB-LD effect. A hallmark of this effect is that the characteristic symmetries of an ideal linear diattenuating retarder Mueller matrix, such as \( M_{12} = M_{21} \), \( M_{13} = M_{31} \), \( M_{23} = M_{32} \), \( M_{24} = -M_{42} \), and \( M_{34} = -M_{43} \) \cite{DuttaGupta2015WaveOptics} are broken as apparent from Fig. \ref{fig:2}(d) (see Equation 6, Supporting information S4). Similar trends were also observed in the corresponding Mueller matrix images \( M(k_x, k_y) \) (Supporting  Fig. S1).  Importantly, as a manifestation of the geometric LB-LD effect, the  \( M_{14} \) (orange dashed box) and the \( M_{41} \) (green dashed box) elements exhibit pronounced spin (circular polarization)-dependent dispersion features. These show spin-dependent intensity  shift  towards \( -\mathbf{k}_x \) for RCP and \( +\mathbf{k}_x \) for LCP, as a  consequence of spin-momentum locking\cite{PhysRevLett.131.193803}.  Note that around the surface plasmon resonance wavelength $636 \ \text{nm}$ (in Fig. \ref{fig:1}(f)),  an additional $\pi$ phase shift is introduced for TM polarization excitation\cite{grigorenko1999phase}. This phase shift reverses the handedness of circular polarization, effecting the conversion of RCP to LCP and vice versa, which in turn results in the sign inversion observed in the $M_{41}$ element (which can be clearly seen seen in the left panel of Fig. \ref{fig:2}(b)).

\begin{figure}
    \centering
    \includegraphics[width=1\linewidth]{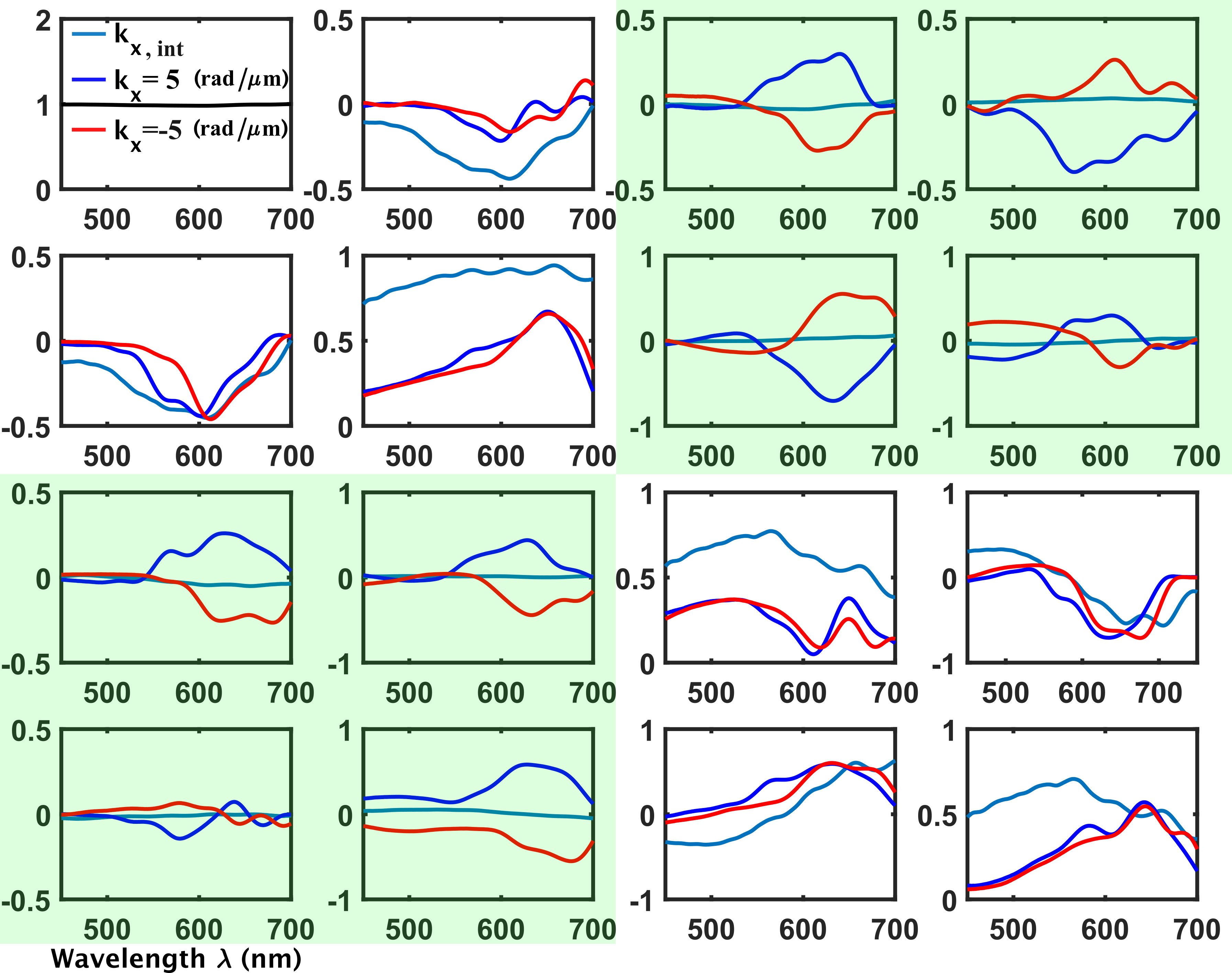}
    \caption{\textbf{Unraveling the geometric origin of the LB-LD effect from the momentum ($\mathbf{k}$)- resolved spectral Mueller matrix of the plasmonic grating:}
    Spectral Mueller matrices are shown for the $k_{x,out}$-integrated response (cyan-blue) and $k_{x,out}$-resolved $k_x = \pm 5 \ rad/ \mu m$ (blue for positive and red for negative $k_{x,out}$). The off-diagonal elements $M_{13/31},\ M_{14/41},\ M_{23/32},\ \text{and}\ M_{24/42}$ reveal the geometric signature of the LB–LD effect through polarity reversal between $+k_{x,out}$ and $-k_{x,out}$. In contrast, the $k_{x,out}$-integrated response eliminates these signatures and subsequently yields near zero magnitudes of the $M_{14/41}$ elements.}
    \label{fig:3}
\end{figure} 

 We now turn to the transverse momentum ($k_{x,out}$)-integrated and $k_{x,out}$-resolved spectral Mueller matrices $M(\lambda)$ (Fig.\ref{fig:3}). The $k_{x,out}$-resolved spectral Mueller matrix elements are obtained by taking horizontal cuts (corresponding to selected $k_{x,out}$) of  \( M(E, \mathbf{k}_{x,\text{out}}) \) presented in Fig. \ref{fig:2}(d), which capture information on polarization transformation due to the trajectory of light as a consequence of tight focusing and  scattering from the plasmonic grating. The ($k_{x,out}$)-integrated Mueller matrix ($M(\lambda)$), on the other hand, is equivalent to space-domain measurement, which encodes information on the intrinsic polarization anisotropy of the plasmonic grating sample. Indeed, the $k_{x,out}$-integrated  $M(\lambda)$ (cyan-blue spectra in Fig.\ref{fig:3}) exhibits the characteristic block-diagonal form \cite{DuttaGupta2015WaveOptics} of a standard linear diattenuating retarder (having magnitudes of linear diattenuation $d_p$ and linear retardance $\delta_p$) with anisotropy axis oriented at $\theta=0^o$ along the grating vector direction.  The corresponding $d_p,\delta_p$ parameters of the plasmonic grating  were estimated using polar decomposition of Mueller matrix (results presented in Supporting Fig. S2) \cite{lu1996interpretation}.
 
 
 The plasmonic grating exhibited strong linear diattenuation due to the preferential excitation of the surface plasmon resonances with TM polarization\cite{ray2017polarization}, with the magnitude of $d_p$ peaking near the resonance wavelength ($\approx 636$nm) (evident from \( M_{12/21} \) elements in Fig.\ref{fig:3}). Considerable magnitude of $\delta_p$ was also observed, which  also attains its maximum value near the resonance wavelength (reflected in \( M_{34/43} \) elements). Notably, the circular anisotropy elements \( M_{14}(\lambda) \) and \( M_{41}(\lambda) \) are considerably low across the entire wavelength range, which is in sharp contrast to the pronounced behavior observed in Fig. \ref{fig:2}(d) (and Supporting Fig. S1).  These results confirm that upon integration over $k_{x,out}$ (or the trajectory of light), the extrinsic geometrical polarization-transformation  is effectively removed, resulting in the elimination of the geometric  LB-LD effect. 

\begin{figure}
    \centering
    \includegraphics[width=0.9\linewidth]{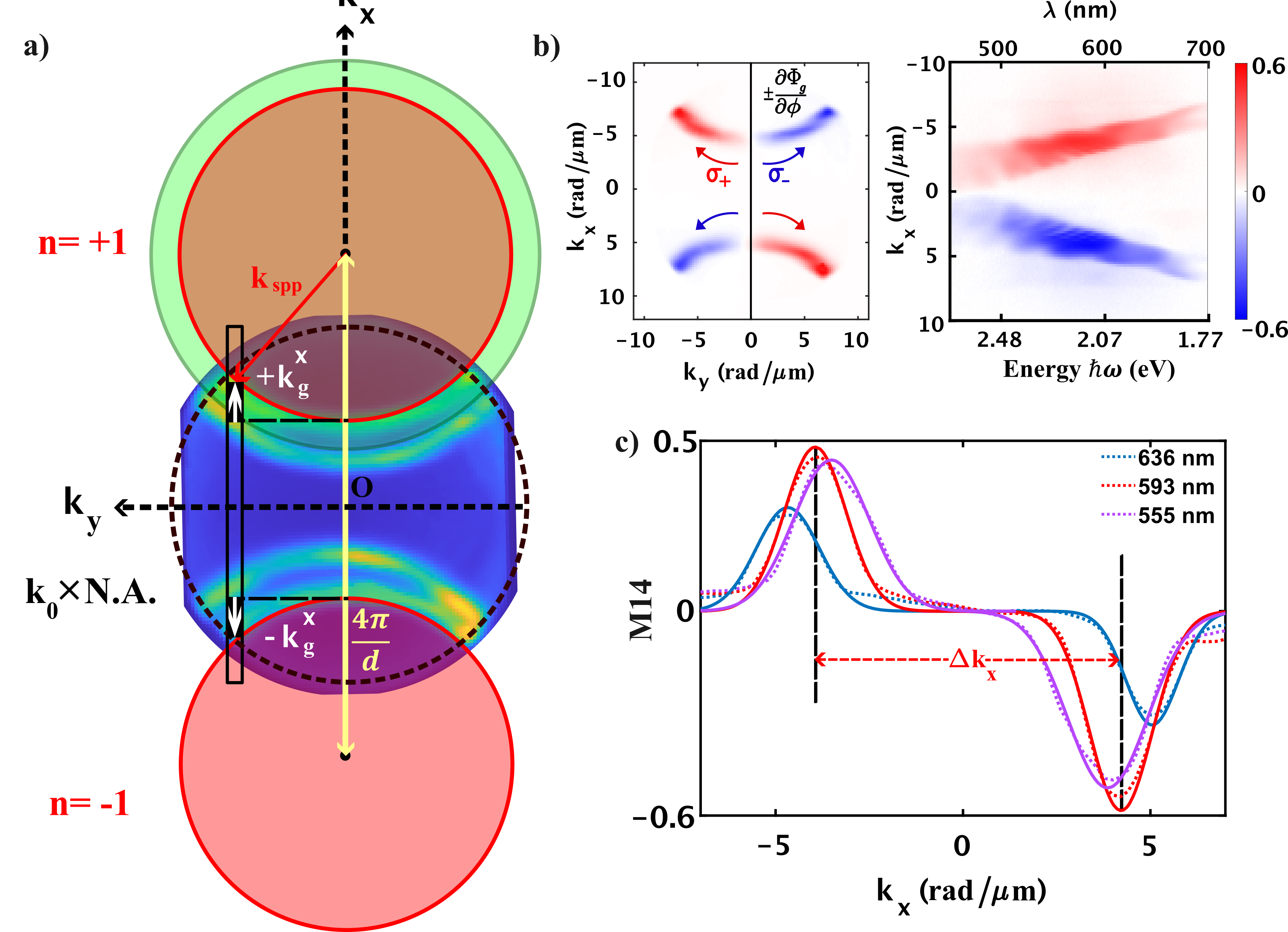}
    \caption{\textbf{Extraction of the spin-dependent transverse momentum ($k_g$) due to geometric LB-LD effect from the experimentally recorded spin-split dispersion Mueller matrix elements $M_{14} (E,k_{x,out})$:} a) Schematic illustration of the construction of the circular overlays corresponding to the Bragg diffraction orders ($-1$ and $+1$). Experimentally recorded arc segments from polarization-blind momentum-domain images are analyzed for two spectral windows: $510–550$ nm (green circle) and $610–635$ nm (red circle). The red circles indicate zeroth ($n=0$) and first-order ($n=\pm1$) diffraction rings for 610–635 nm, while the black dotted circle marks the numerical aperture of the imaging system. (b) Clockwise and counter-clockwise rotation of RCP/ LCP polarized intensity lobes in the momentum domain \( M_{14}(k_{x}, k_{y}) \) element (left panel). The corresponding dispersion of the Mueller matrix element $M_{14}$ as a function of transverse momentum $k_x$ and photon energy $E$ (or wavelength $\lambda$), showing spin-dependent splitting.}
    \label{fig:4}
\end{figure}

The $k_{x,out}$-resolved spectral Mueller matrices (shown for $k_{x,out} =\pm5 \ rad/\mu m$  by blue and red colors respectively, in Fig. \ref{fig:3}), on the other hand, show pronounced dependence of the \( M_{14/41} \) elements on $k_{x,out}$, underscoring the geometric nature of the LB-LD effect. Moreover,  the usual symmetry and the block-diagonal form of the  Mueller matrix is altered with significant intensities of the off-block diagonal elements ($M_{13/31},\ M_{14/41},\ M_{24/42},M_{23/32}\ \text{and}\ M_{24/42}$) (marked in Fig.\ref{fig:3})\cite{DuttaGupta2015WaveOptics,bliokh2015spin}. These elements bear exclusive signatures of the underlying geometrodynamics of polarized light of the geometric LB-LD effect. As predicted by our model of geometric LB-LD effect (Eq. 4, Supporting information S4), the off-block diagonal elements including the \( M_{14/41} \) elements exhibit polarity reversal between opposite transverse momenta $+ k_{x,out} \ \text{and} - k_{x,out} $. This can be understood by noting that the two sequential polarization transformations (tight focusing and scattering from the anisotropic plasmonic grating, as previously discussed) lead to light trajectory-dependent azimuthal (\( \phi \)) variation of the relative orientation of the anisotropy axis of the two individual effects in the momentum domain. Since, opposite azimuthal trajectory of light $( \pm  \phi) $ are associated with opposite transverse momentum components $\pm k_{x,out}$, the corresponding polarization effect is reflected as polarity reversal in the $k_{x,out}$-dependent Mueller matrix elements.  These features are also apparent in the Mueller matrix-derived orientation angles of both the linear diattenuator and the linear retarder (see Supporting Figure S3)  Together, these results provide conclusive evidence of the geometric nature of the LB-LD effect. This effect generates a geometric phase gradient which does not arise from the structural anisotropy gradient of the system (the inversion symmetric plasmonic grating) but due to the interaction of the structured polarization of focused light with the anisotropic plasmonic grating. The resulting geometric phase gradient provides an additional spin-dependent transverse momentum ($\pm k_g$) which leads to the unconventional spin-split dispersion in the inversion-symmetric system ($\mathbf{k_g}=\sigma\nabla\Phi_g,\sigma=\pm1$ for RCP and LCP)\cite{rong2020photonic}. In what follows, we determine the additional spin-dependent transverse momentum ($k_g$) from the recorded spin-split dispersion Mueller matrix elements $M_{14} (E,k_{x,out})$ (shown in \ref{fig:4}).

\subsection*{\textit{Extraction of the spin-dependent transverse momentum ($k_g$) arising due to geometric LB-LD effect:}}
The four-fold azimuthal intensity lobes in the circular polarization-resolved intensity arc segments of the recorded momentum domain Mueller matrix element \( M_{14}(k_{x}, k_{y}) \)  (\ref{fig:4}b) provide evidence of azimuthal geometric phase evolution   $~\exp^{\pm \iota m \phi}$, with m = ($\partial \Phi_g /\partial \phi$)=1. This follows from the fact that SOI in vortex (azimuthally varying) diattenuating retarder (inhomogeneous anisotropic medium) system is generally characterized by $~ 4m$-fold spin-polarized azimuthal intensity lobes\cite{bliokh2015spin, DuttaGupta2015WaveOptics}.  The opposite azimuthal geometric phase gradient ($\pm \partial \Phi_g /\partial \phi$) for RCP/ LCP polarization states imparts clockwise and counter-clockwise rotations (shown by red and blue arrows in \ref{fig:4}b) of the RCP/LCP (\(\sigma = \pm 1\)) polarized intensity lobes in the momentum \( (k_{x}, k_{y}) \)plane \cite{DuttaGupta2015WaveOptics}. This provides the additional spin-dependent transverse momentum ($\pm k_g^{x}$) in the momentum matching condition of the leaky surface plasmons $k_{x,out} = k_{spp}-2\pi/d \pm  k_g^{x}$ for the exhibition the spin-split dispersion in our inversion symmetric plasmonic grating system (as shown by the white arrow in \ref{fig:4}a). In order to estimate ($\pm k_g^{x}$) from the momentum domain image of the Mueller matrix elements \( M_{14}(k_{x}, k_{y}) \) and from the corresponding spin-split dispersion $M_{14} (E,k_{x,out})$, we proceed as follows. The $k$-domain images for two distinct spectral regions ($ 510–550\ nm $ and $610–635\ nm $) were spatially integrated. The resulting image, shown in Fig. \ref{fig:4}(a), exhibits two distinct arc-like diffraction patterns corresponding to each spectral region. These arc segments are then used to construct circular overlays corresponding to the Bragg diffraction orders ($-1$ and $+1$), as shown by the red (for $\lambda=610-635\ \text{nm}$) and green circles (for $\lambda=510-550\ \text{nm}$) for both spectral regions. The magnitude of $k_g^{x}$ is subsequently extracted from the above-mentioned momentum matching condition. For this purpose, the differential  transverse momentum $\Delta k_{x,out}$ between the LCP and the RCP states were extracted from the experimentally recorded dispersion $M_{14} (E,k_{x,out})$ (shown in \ref{fig:4}c) for the three selected plasmon resonance wavelengths 636 nm (blue), 593 nm (red), and 555 nm (magenta). The magnitude of $k_g^{x}$ was estimated to be $\approx 1.2\ rad/\mu m$, which is quite strong and found to be relatively insensitive to the wavelength.

\section{Conclusion} 

To summarize, we have observed an extraordinary spin- dependent dispersion effect of leaky surface plasmons in one dimensional plasmonic crystal grating exhibiting perfect spatial inversion symmetry. Momentum-domain polarization  measurements unraveled that the observed spin-split dispersion arises from a transverse momentum (light trajectory)-dependent interplay of linear birefringence and linear dichroism effect, the so-called geometric LB–LD effect. This effect is a geometrical counterpart of the recently reported "intrinsic" LB-LD effect that leads to giant chirality in centrosymmetric crystal \cite{Parrish2025CentrosymmetricCD}. Analysis of the momentum ($\mathbf{k}$)-resolved spectral Mueller matrices and  Mueller matrix dispersion  $M(E,k)$ of the plasmonic crystal  provided valuable insights on this unconventional SOI of light mediated by the geometric LB-LD effect. The two-step process of the geometric LB-LD effect involving polarization evolution due to the focusing transformation and subsequent interaction of the structured polarization field with the inversion-symmetric anisotropic plasmonic grating system leads to the emergence of strong  geometrical phase gradient or spin (circular polarization)-dependent transverse momentum, which in turn governs the spin-dependent splitting of the leaky surface plasmon modes.  We have  quantified this additional spin-dependent transverse momentum from the experimental  spin-split dispersion by applying momentum-matching condition  for  surface plasmon excitation and out-coupling of the leakage radiation. Our findings on spin-dependent dispersion in a simple symmetric system delves into new realm of spin-based dispersion engineering and opens exciting opportunities for spin-controlled nano-optical functionalities using the interplay of structured optical field polarization and metasurfaces through the geometric LB-LD effect.



\bibliography{sn-bibliography}

\end{document}